\documentclass[journal=ancac3,manuscript=article]{achemso}

\usepackage[T1]{fontenc}       
\usepackage{amsmath}
\usepackage{subfig}

\author{Niels R. Walet}
\email{Niels.Walet@manchester.ac.uk}
\affiliation{Theoretical Physics Division, School of Physics and Astronomy, University
of Manchester, M13 9PL, UK}
\title{Water confined between graphene layers: the case for a square ice}

\begin{document}

\begin{tocentry}
\includegraphics[width=8.5cm]{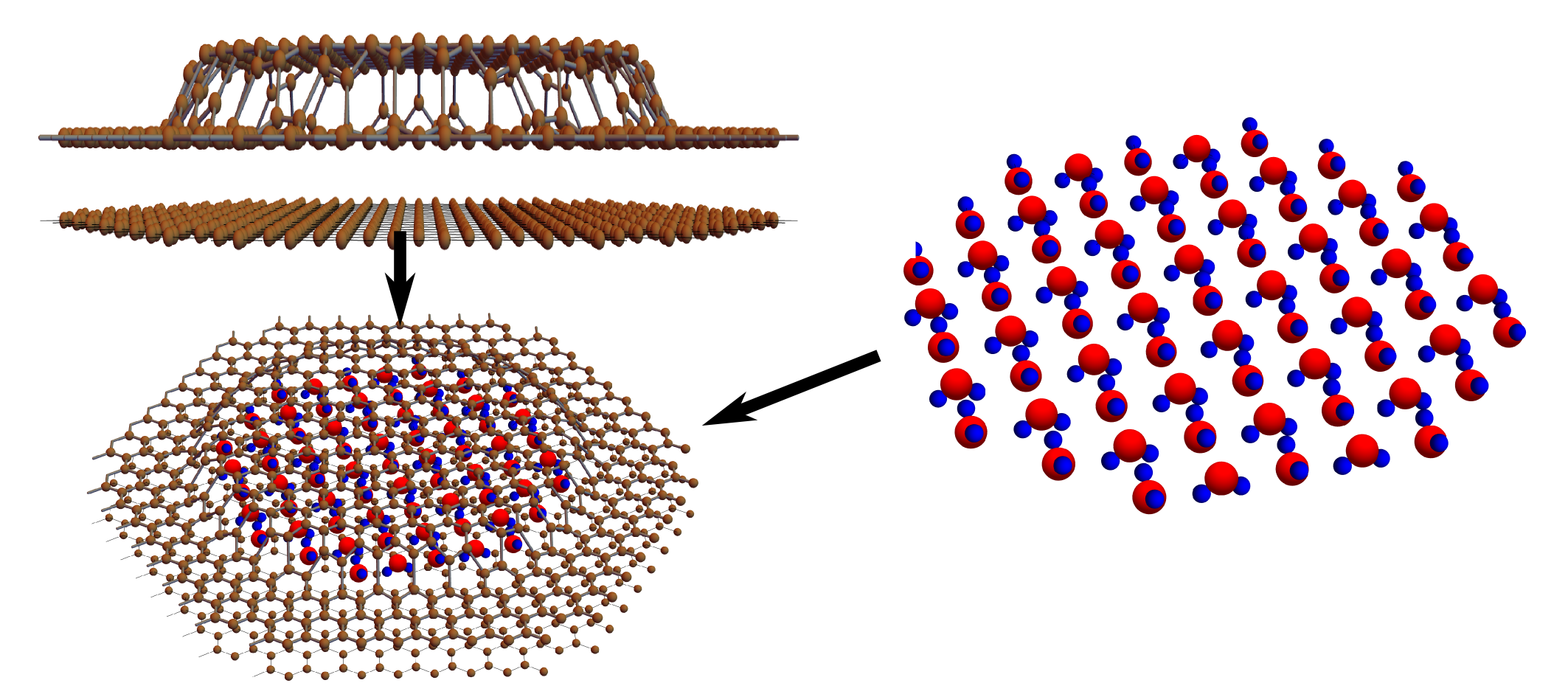}
\end{tocentry}

\begin{abstract}
  Water confined between two graphene layers with a small separation
  forms a two-dimensional ice structure,with an apparent square
  symmetry {[}Algara-Siller \emph{et al.}, Nature (London) \textbf{519}, 443
  (2015){]}, which is poorly understood.  
  A density functional approach is applied to the water, but not to the water graphene interactions, since the two crystals are incommensurate.
 We thus need use a potential model for the interaction between water and graphene.
  We analyze the
  models for confinement of water by graphene, and find that even
  though  the general features are well established, the detail is not so
  well understood. Using  a representative range of potential models,
  we perform density-functional calculations and show that many ice-like configurations exist. 
  In some cases these are unstable with respect to decay into
  a bi-layer structure, but we expect tunneling between such structures to be slow.
  It is shown that there is one good candidate
  for a square crystal, which is a peculiar anti-ferroelectric
  arrangement of water molecules. 
\end{abstract}


Considering how important water is in our daily life, it should come
as no surprise that understanding its behavior in all kinds of situations
is of great interest, even in rather special ones. One recent example
is water confined at the nanoscale, which has been studied extensively
theoretically\cite{fraerman1999metastable,kolesnikov2004anomalously,kumar2007effectof,cicero2008waterconfined,mogelhoj2011abinitio,kirov2012newtwodimensional,kirov2013residual,anick2014staticdensity,ramirez2014iceand,zhao2014ferroelectric,akashi2015firstprinciples,chen20152dice,standop2015h2oon,corsetti2016structural,sobrinofernandezmario2015aastacked},
and received a considerable boost following the discovery of a square
crystal when water is confined between two layers of graphene \cite{algara-siller2015squareice}.

The experiments seem to show a perfect square crystal structure,
and indications are that this is not a bilayer structure
where one layers is offset relative to the other (so-called AB stacking).
Bilayer structures are also seen in the experiment, and
claims are made that this is consistent with AA stacking \cite{sobrinofernandezmario2015aastacked}.
The oxygen-oxygen distance is quoted as $2.8\,\mathrm{\mathring{A}}$.
In the paper by Algara-Siller \emph{et al.} \cite{algara-siller2015squareice} a classical molecular
dynamics calculation is also reported, that shows some aspects of agreement
with the experiment, but seems to indicate a rhombic crystal. Such
results should be taken with a grain of salt, since potential models
are notoriously unreliable for water \cite{sokhan2015signature}. This
is one of the reasons why so many potential models exist
\cite{mahoney2000afivesite,ren2003polarizable,horn2004development,rick2004areoptimization,abascal2005ageneral,fanourgakis2006theflexible,fanourgakis2008development}.
One of the problems of all these models is a poor description of the
higher electric multipole moments (beyond dipole), which are expected
to be important in the behaviour of water \cite{abascal2007themelting}. 
Indeed, the results of the MD simulations 
\cite{algara-siller2015squareice}
show a rhombic ice, where thermal motion needs to be invoked to show
a structure that is square on average, which feels slightly artificial,
but may be possible. 

To avoid the uncertainty of potential models a number of investigations
have used density functional theory to describe the interaction of
the water molecules, employing simple confining potentials to account
for the effect of graphene. Using such an approach one can only study
the zero-temperature ground state, which may be a draw-back, unless
this structure is energetically well separated from other states.
We are aware of one study \cite{partovi-azar2015vander} that looks
at a water layer in front of a single layer of graphene using DFT
at finite temperature, but none that look at the problem at hand.
On the other hand there are a number of fully microscopic calculations
of water in front of finite but large flat carbon molecules have been
performed \cite{cabaleiro-lago2009studyof,jenness2009dfdftsapt,rubevs2009structure,jenness2010benchmark,kysilka2011accurate,voloshina2011onthe}.
These have been extrapolated to graphene, and probably give a decent
estimate of the binding and orientational dependence of the binding
of water to graphene. The long-range behavior of all such models
is $r^{-6}$, showing the fact that the large but finite carbon molecules are all
insulators--unlike real graphene, which is a semi-metal. Nevertheless, we shall use
these results to build improved models. There are many ways to model the interaction of water with graphene. We shall use a simple surface-integrated
Lennard-Jones potential (3-9 potential) which has the correct assymptotics \cite{guinea2016interaction},
but a rather arbitrary  choice of core repulsion (see below). 

One of the other problems that all investigations face is the choice
of thermodynamic potential to be optimized when describing confined
water--this is a thorny question, and most papers adopt a form of
2D enthalpy linked to the inter-layer separation distance and the
in-plane stress (rather than the 3D pressure). There is no obvious
reason to look at constant stress; the real situation is much more
complicated than in an ideal 2D baryostat, since the vertical stresses
must have an influence on the separation of the graphene layers. 

We approach the problem in a slightly different way: we report a set
of DFT calculation of a model system at a density corresponding exactly
to the experimental one (one $\mathrm{H}_{2}\mathrm{O}$ molecule
per $(2.8\,\mathrm{\mathring{A}})^{2}$). The picture that underlies this idea is
that the graphene response to stresses is rather complex, with layers
separating rather suddenly if a pressure/stress threshold is reached.
As long as stresses are relatively small, that is probably a reasonable
approach to take; when the vertical stress increases, we expect the
graphene layers to separate further and accommodate multiple layers.
This is actually an example of a well-established mechanism,
which is used in the production of intercalated graphite compounds\cite{ebert1976intercalation}. 

We will first, in the next
section, investigate the nature of the confinement water by graphene and the interaction of
water with graphene, including how the interaction depends on the orientation of the water molecule.
Then we analyze in some detail, in the next section, DFT calculations
and the dependence on the confining potential. In the final section
we give look at conclusions and potential further work.

\section{Confinement of water by graphene}

\subsection{Water-graphene interaction potential}

We first investigate the nature of the confining potential. The interaction
of water with graphene has been studied microscopically, using coupled
cluster theory \cite{voloshina2011onthe,cabaleiro-lago2009studyof},
many-body perturbation theory \cite{jenness2009dfdftsapt,jenness2010benchmark}
and a DFT/perturbation theory hybrid \cite{rubevs2009structure,kysilka2011accurate}.
These results set a benchmark: a maximal attraction of in the order
of $100-130\text{ meV}$ per molecule at a distance of about $3.2\text{Å}$.
The detailed data on distance dependence in figure~2 of 
Voloshina \emph{et al.}
\cite{voloshina2011onthe}
seems to show an attraction that falls as $r^{-6}$,
which appears inconsistent with the semi-metallic nature of graphene,
and probably reflects a limitation of the extrapolation to graphene of a long-range interaction
of water with a finite carbon molecule, which unlike graphene is insulating. It also
shows, in the region where the calculations must be more reliable, since
it is dominated by electronic correlations, a soft repulsive core
that is poorly fitted by an inverse power law. Most importantly for
the work reported here, the harmonic part of the potential about equilibrium
is rather small, and much less than estimated from a van-der-Waals potential ($3-9$) or $6-12$).
The attraction between water and graphene depends on the water orientation
and its position relative to the graphene layer, but this is not the case
for the equilibrium distance  of the molecule relevant to graphene. 
This can be compared to the finite-temperature
DFT (and thus infinite through the use of periodic boundary conditions)
calculations in \cite{partovi-azar2015vander}, which suggest that
a randomly oriented water slab sitting in front of a single graphene
layer feels a potential (lowest curve in their figure~4) that can be
fitted by an $1/r^{3}$ long range attractive force, again with a
weak repulsive kernel that does not seem to follow a simple
power law--definitely not the strong repulsion used in most works
on the subject. \footnote{Part of this discrepancy may be due to their definition of separation as
the distance of the nearest hydrogen atom to the graphene layer, which
is inconsistent with the more common definition of the distance from
the oxygen molecule or almost equivalently the center-of-mass to the
graphene layer. The distance quoted is therefore always smaller than
the usual definition of distance, since, as the authors state, the
Hydrogen bonds prefer to point towards the graphene layer.} 

For our calculation, based on the experiment where the ice crystals
are incommensurate with graphene, we can not describe the graphene
and water simultaneously in full microscopic detail, and we need a
potential model to describe part of it--as usual we do this for the
graphene-water interaction. There are a number of models for the interaction
between a material such as graphene and water; some (e.g. \cite{corsetti2016structural})
use the $3-9$ hydrophobic interaction first proposed 
by Lee \emph{et al.}
\cite{lee1984thestructure};
Chen \emph{et al.} \cite{chen2016twodimensional} use a Morse potential
fit to a rather scattered set of GFMC data, and modern potential models
for water in carbon-nanotubes (see, e.g., \cite{jaffe2004watertextendashcarbon,perez-hernandez2013anisotropy,kaukonen2012lennardjones,hummer2001waterconduction,werder2003onthe})
suggests a fit of the interaction between water and carbon atoms as
a combination of C-O and C-H $6-12$ Lennard-Jones potentials. 

Actually
we only need a full model of the potential for our discussion of graphene
bubbles below, whereas in the rest of the paper it will suffice to
look at a harmonic expansion. For the analysis below we mainly use
the simplest model, an interaction between the graphene and the oxygen
in the water molecule only. In the light of the discussion above,
we choose a $3-9$ potential of the form 
\begin{equation}
V_{39}(z)=V_{0}\left[\frac{1}{2}\left(\frac{\sigma}{z}\right)^{9}-\frac{3}{2}\left(\frac{\sigma}{z}\right)^{3}\right],\label{eq:39}
\end{equation}
where for the water-graphene potential, $V_{39}^{wg}(z)$, we use
\begin{equation}
V_{0,wg}=120\text{ meV},\ \sigma_{wg}=3.2\text{Å}.
\end{equation}
The strength used here is just a suitable average of the microscopic
values, which range from $-100$ to $-135\,\text{meV/molecule}$,
but do depend on position and orientation relative to the graphene
layer. The equilibrium position is a similar average, but is not very
sensitive to position or orientation \cite{voloshina2011onthe}.

One simple way to include the directionality of the water-graphene
interaction is to fit two 3-9 potentials (\ref{eq:39}) to the microscopic
data, a graphene-O and a graphene-H one. For all the reasons discussed
above, this has an unreasonably large harmonic behavior and also
lacks the dipole--semi-metal long-range interaction \cite{guinea2016interaction},
but with limited data available this is probably the best we can do.
We fit the results \cite{voloshina2011onthe} with a form
\begin{equation}
V(z,\theta)=V_{39}^{Og}(z)+V_{39}^{Hg}(z+b cos(\phi_b/2+\theta)) +V_{39}^{Hg}(z+b cos(-\phi_b/2+\theta)),
\end{equation}
where $\theta$ is the angle of the dipole moment with the normal, assuming that the Hydrogen atoms are always in a plane perpendicular to the graphene layer.
We find that the strength parameters are 
\begin{equation}
V_{0,Og}=77.7\text{ meV},\,\sigma_{Og}=3.14\text{ }\text{Å},\,V_{0,Hg}=24.3\text{ meV},\,\sigma_{Hg}=2.73\text{Å},\label{eq:VHVO}
\end{equation}
which reflects the fact that for a single layer the dipole moment points preferentially towards the graphene  layer.

\subsection{Graphene bubbles}

Graphene bubbles are rather difficult to model, because there are
many different situations we can consider. 
It seems likely that for
thin layers of water confined between graphene, the model of Guinea
[F. Guinea, Pressure inside crystalline bubbles confined by graphene, 2016]
applies. This models an extended bubble with height
much smaller than its width, which is flat in the middle, and has
a smooth connection to the underlying substrate of width $d$. Of
course more complex models can be constructed; the best type of mechanical
model is probably one based on thin-plate theory (see, e.g., \cite{wang2013numerical}),
but even that has weaknesses, since adhesion is quite a complex phenomenon, as
for instance discussed 
by Sprigman and collaborators
\cite{springman2008snaptransitions}.
We shall ignore all such complexity, and use the simple model, since
here we are more interested in the phase inside the bubble than the
exact details of the shape and dynamics of the bubble. 

\begin{figure}
\begin{centering}
\includegraphics[width=7cm]{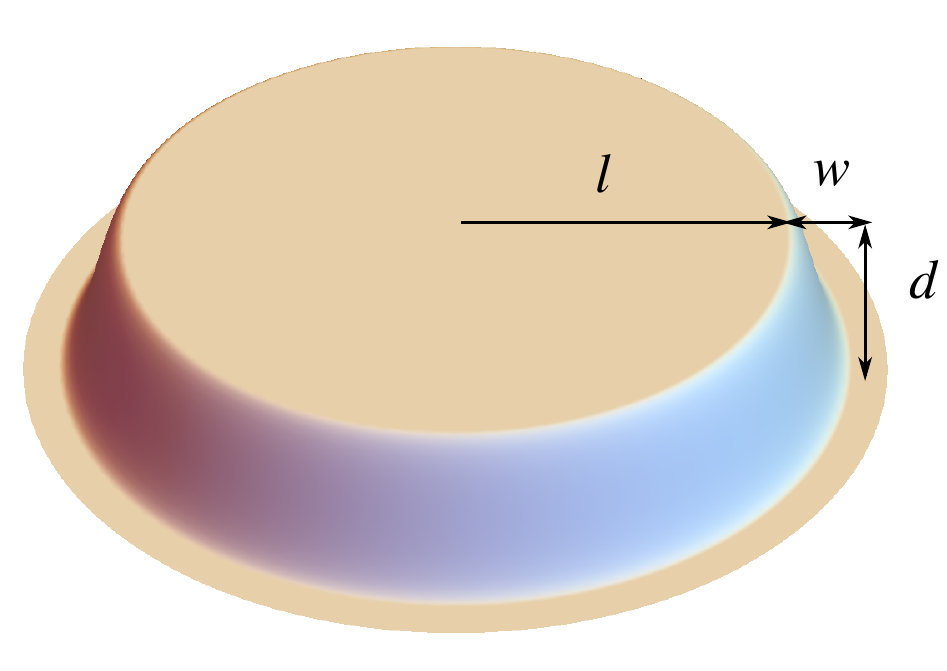}
\par\end{centering}

\caption{A graphical representation of the confinement
model. We assume that the boundary is sharp, $w\ll l$, and that the
height $d$ is constant, apart from the boundary region, with $d\ll l$.\label{fig:pillbox}}
\end{figure}

Many discussions use ``pressure'' as the quantity describing the
confined phase. This is not correct: it is straightforward to show
that the stress in the direction perpendicular to the graphene is
very different from that in the two directions parallel to the layers.
If we assume a cylindrical bubble, it seems reasonable to assume the
in-plane stress is diagonal and equal in both directions, and this
may be what other work calls the ``pressure''. As we shall see the
in-plane stresses range from $0.1\text{ GPa}$ upwards, whereas the
vertical stress, to be defined below, at the experimental density
should be small. Stresses are calculated as the variation of energy
with the dimensions of the simulation box, which lacks some of the
detailed information that might be of interest, since the horizontal
force for a layered system is most likely to be strongest at the height
of the confined layer(s), and almost zero near the graphene layers. Nevertheless,
with the data available we can construct a crude model of how graphene
confines the water. We first ignore the internal elasticity of graphene,
and assume the water to be confined inside a pill-box shaped graphene
container of radius $l$, see figure~\ref{fig:pillbox}. If we assume
that the area of bubble is much larger than the graphene spacing,
we can determine the equilibrium distance ignoring the sides, by minimizing
the potential energy per unit area. 

We model this by a total of three $3-9$ potentials; two representing
the interaction between two layers of water (assumed at a distance
$(d\pm d_{1})/2$ of each of the two graphene layers) and each of
the two graphene layers, and one for the attraction between the two
graphene layers at distance $d$, 
\begin{equation}
V(d,d_{1})=V_{39}^{wg}((d+d_{1})/2)+V_{39}^{wg}((d-d_{1})/2)+V_{39}^{gg}(d).\label{eq:totaV}
\end{equation}
Here the water-graphene potential $V^{wg}$ was already introduced above and 
we denote by $V^{gg}(d)$ the potential energy per unit area for two
graphene layers in vacuum at separation $d$. If we take the GFMC calculations
\cite{mostaani2015quantum} as a benchmark, we find that a reasonable
value for $V_{0,gg}=-V_{39}^{gg}(d_{0})$ is about $8\text{ meV/\ensuremath{\text{Å}^{2}}}$,
where the equilibrium distance $d_{0}=\sigma_{gg}=3.41\ \text{Å}$.
For water, we use the parameters defined above, $V_{0,wg}=120/2.8^{2}=15.3\text{ meV/\ensuremath{\text{Å}^{2}}}$. 
If we ignore the edges of the bubble, i.e., $l\gg d,w$, we can
minimize the energy per unit area Eq.~(\ref{eq:totaV}) to find that
the optimal configuration is a single layer of water in the middle
of two graphene layers separated by $d=6.36\ \text{Å}$, which is
very close to $2\sigma_{wg}$, since the water-graphene potentials
are stronger than the graphene-graphene one. This also means that
this result is essentially independent of the potential model, since
it is mainly sensitive to the equilibrium distance and binding energy. 

For generality, we shall look at the  energy associated with the classical potentials 
of a bilayer of water, a distance $(d\pm d_1)/2$ from the graphene layers, 
\begin{equation}
E=\pi l^{2}\Delta
V^{gg}(d)+\pi l_0^2 \left(V^{wg}((d+d_1)/2)+V^{wg}((d-d_1)/2)\right).\label{eq:vol}
\end{equation} 
Here $\Delta
V^{gg}(d)=V_{39}^{gg}(d)-V_{39}^{gg}(d_{0})$, and $l_0$ is the length
parameter where we have evaluated the water-graphene interaction parameters; since the
number of water molecules is fixed, in reality the energy depends on
the numbers of water molecules, not on the size of the bubble.
The horizontal force exerted by the graphene on the water due to a
change in the radius $l$ is thus $F=2\pi l\Delta V^{gg}$. If the
in-plane stress tensor of the water is diagonal with equal magnitude,
$S_{x}=S_{y}=S$, we see that the average force on the rim of the
bubble is $2\pi l\,\Delta d\,S$, which must balance the force exerted
by the bubble due to the interactions between the water molecules.
 Comparing the two expressions for the force, we find
the relation
\begin{equation}
S=\Delta V^{gg}(d)/\Delta d.
\end{equation}
This simply states that the stress of confinement balances the horizontal
stress exerted by the confined water. If we use $d=6.4\text{Å}$,
we find $\Delta V^{gg}\approx6\text{ meV}/\text{Å}^{2}$
which is
somewhat dependent on the potential model. This then gives a value
for $S=S_{x,y}$ of the order of $S=0.5\text{ GPa}$, which is rather similar to
the numbers found experimentally by Vasu \emph{et al.} \cite{vasu2016vander}, which are maybe a factor of 2 higher. Our
estimate could potentially vary by factors of 2 up or down even if
we were to know $\Delta V^{gg}$ more accurately: As shown by Guinea
the elastic energy of graphene tends to increase
the force exerted on the water, whereas the expected inhomogeneity
of the force exerted by the layer of water tends to induce a stronger
curvature of the side of the graphene, thus supporting a lower stress.
This analysis is only valid in the absence of vertical stress. 
It is quite close to the  values of "pressure" reported
by Vasu \emph{et al} \cite{vasu2016vander}.

\begin{figure}
\includegraphics[width=8cm]{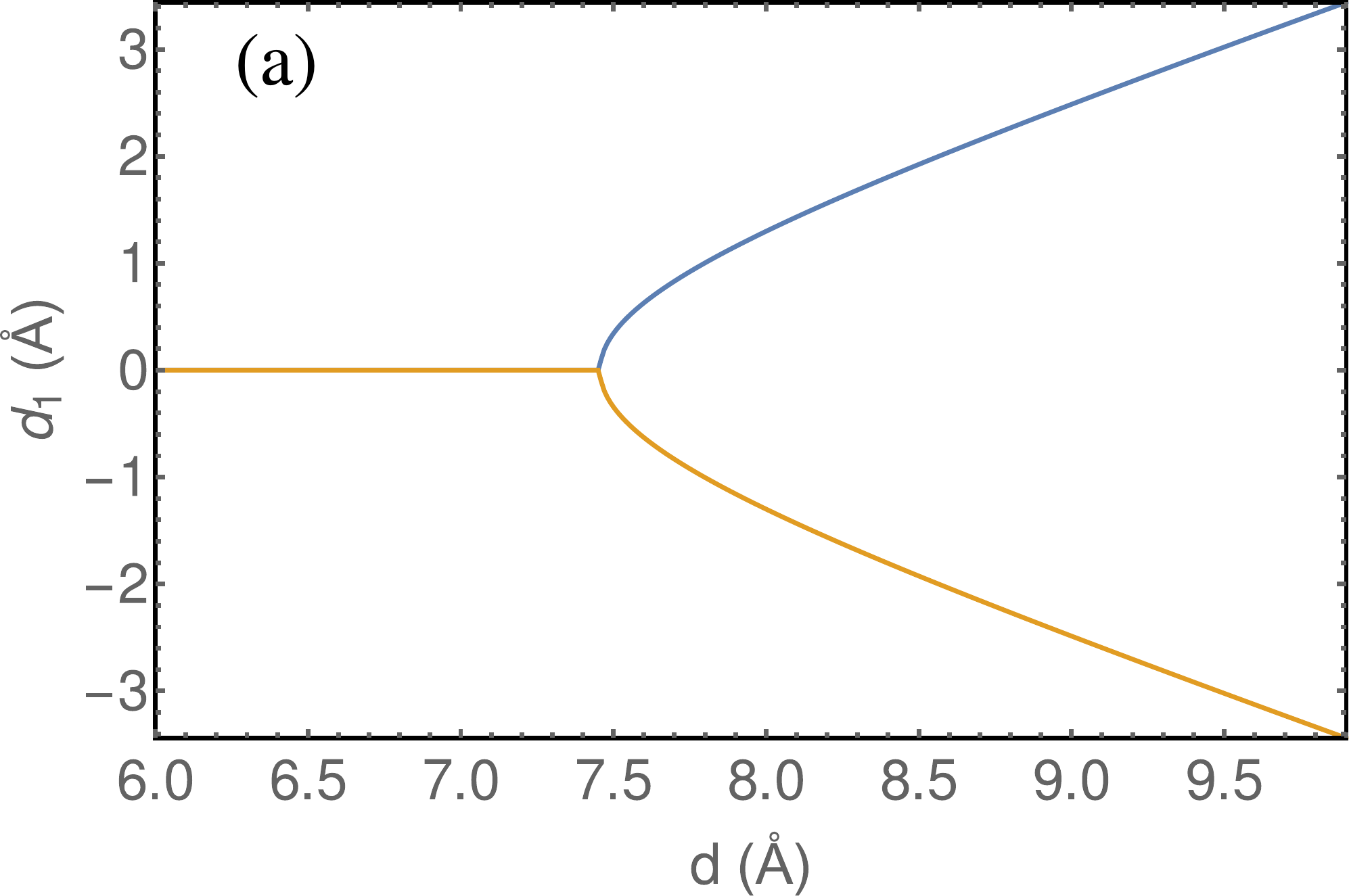}\quad\includegraphics[width=8cm]{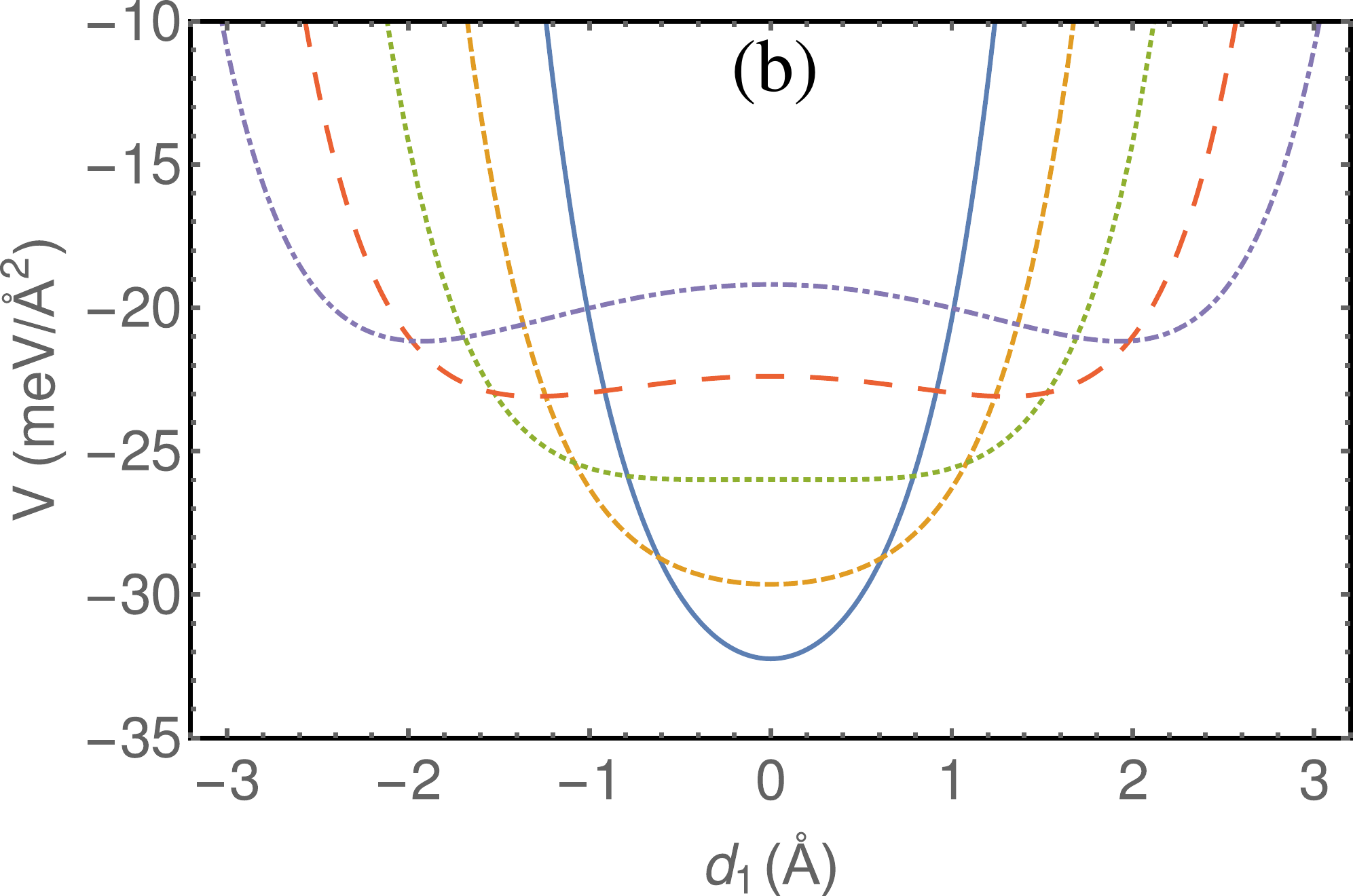}
\caption{ Analysis of the potential energy per unit area
$V,$ Eq.~(\ref{eq:totaV}) as a function of $d$ and $d_{1}$.\label{fig:d1d}
(a) The optimal choice of the water-water distance $d_{1}$ as a function
of the graphene-graphene distance $d$. At $d=7.45\ \text{Å}$ bilayer
forms. (b): 
The shape of the potential $V$ as a function of $d_{1}$ for $d=6.5\text{Å}$
(solid blue line), $d=7\text{Å}$ (yellow dashed line), $d=7.5\text{Å}$
(red dotted line), $d=8\text{Å}$ (green long-dashed line) and $d=8.5\text{Å}$
(purple dot-dashed line).
}
\end{figure}

If we artificially increase the separation between the layers, i.e.,
look at the optimal values of $d_{1}$ for fixed $d$, see figure~\ref{fig:d1d},
we find that the potential becomes shallow quite quickly. As we force
the graphene layers further apart, there is a sudden transition from
a single water layer to a bilayer--for our simple model this bifurcation
occurs at $d=7.45\ \text{Å}$. 

Since the water itself also exerts a vertical
stress, the picture changes: as we increase the density both horizontal
and vertical stress will increase, leading either to an increase in
radius of the bubble, or an increase in height with the concomitant
formation of a bilayer, which may induce some of the snap transitions discussed
in the literature
\cite{springman2008snaptransitions}. 
The vertical force exerted by the graphene can be calculated as $F=\pi l^{2}\partial_{d}V^{gg}$.
If $d$ is almost the equilibrium distance as shown above, this is
almost zero, and thus the vertical stress is almost zero as well.
When the stress of the water increases, we see that we must move away
from this equilibrium. At the same time the reaction force to the
confining potential also contributes, and we need to balance stress
from the water, the attraction exerted by the water due to confinement,
and the graphene-graphene interaction. If we use the potential model
sketched above, we find this is mainly sensitive to the harmonic part
of the water-graphene potential. If we use the $3-9$ potentials as
above (which leads to too large a curvature of the potential), we find that the graphene
layer moves roughly $0.3\text{Å}/\text{GPa}$; this is clearly strongly dependent 
on the potential model, but suggests that realistic
numbers are less than 10 times that value, which is still a small effect, if we realize that stresses are typically less than $1\,\text{GPa}$.

\subsection{Orientational order}

\begin{figure}
\begin{centering}
\includegraphics[width=8cm]{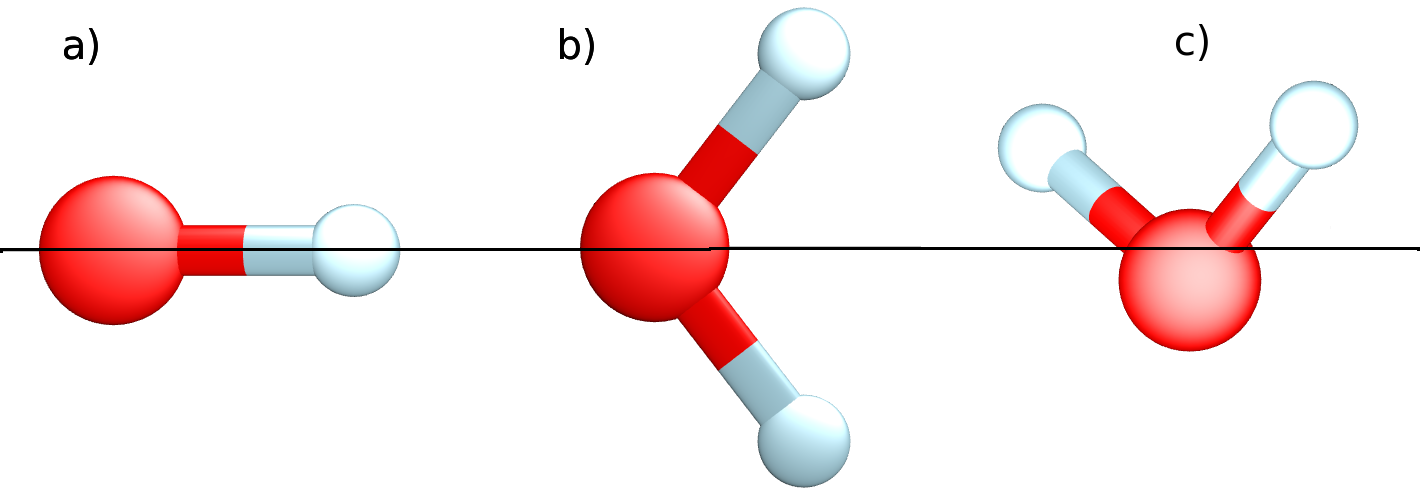}
\par\end{centering}

\caption{The three positions of a water molecule considered
for interactions, when confined between two metallic layers at distance
$5\text{Å}$ parallel to the black line, and perpendicular to the
paper. The black line gives the mid-point between the layers, and
the positions of the atoms are the ones obtained after optimization
(most important for case c).\label{fig:The-three-positions}}
\end{figure}

\begin{table}
\centering{}\caption{The energy gain of a water molecule placed between two metallic layers,
as compared to the same molecule in vacuum, see figure~\ref{fig:The-three-positions}
for details.\label{tab:The-energy-gain}}
\begin{tabular}{|c|c|}
\hline 
case & energy gain\tabularnewline
\hline 
\hline 
a & -12.5 meV\tabularnewline
\hline 
b & -19.2 meV\tabularnewline
\hline 
c & -57.5 meV\tabularnewline
\hline 
\end{tabular}
\end{table}

We now look at the orientation dependence of the energy of a  water molecule
between 2 graphene
layers, which is an important secondary aspect of the problem. We
shall use input from microscopic calculations using water in position
c) in figure~\ref{fig:The-three-positions} relative to a \emph{single}
layer of graphene. For the case of a bilayer, it is relatively straightforward
to show \cite{guinea2016interaction} that the interaction of a point
dipole between two graphene layers has a preferential orientation.
In the RPA approximation for graphene, if the dipole makes an angle
$\theta$ with respect to the vertical to the layers, we can find
an analytical expression \cite{guinea2016interaction}. Using this
with $d=5\text{Å}$, and $p=1.85\text{ D}$ for water, we then find
$V=-(31+51\cos^{2}\theta)\text{ meV}$ for a metal ($\epsilon_{r}=\infty)$,
which gets reduced to $V=-(21+29\cos^{2}\theta)\text{ meV}$ for two
free-standing graphene layers with the estimate $\epsilon_{r}=5$.
This is entirely consistent with the variation seen in the microscopic
calculations cited above (but there one normally only considers a single
layer with the two realizations of case c, with the dipole moment pointing
to or away from graphene), suggesting this is a relevant effect.

We have also calculated these numbers for a slightly more realistic description
of water, using a slightly augmented version of the approach as used
below for ice: we use the EMS method \cite{otani2006firstprinciples}
within a DFT calculation using quantum espresso \cite{giannozzi2009quantum},
which makes it possible to add electrostatic boundary conditions and
remove the usual 3D periodicity on the direct interactions, to study
a single water molecule between two perfectly structureless metallic
layers ($\epsilon_{r}=\infty$)\footnote{Even though the case of finite $\epsilon_{r}$ is discussed 
by Ottani \emph{et al.}
\cite{otani2006firstprinciples},
this case is not actually implemented in the quantum espresso code,
since it requires an additional model-dependent solver step, where
a charge distribution must be inferred [M.~Otani, private communication]}, separated by $5\text{Å}$ to enhance the effect of the boundaries.
We compare the energy difference between for same molecule between
two metallic layers ($\epsilon_{r}=\infty$), and two layers of vacuum
($\epsilon_{r}=0$) for the three configurations shown in figure~\ref{fig:The-three-positions}{]},
with a structural optimization that keeps the orientation of the water
molecule. The interesting aspect is the difference between cases a)
and b), which is not analyzed in detail in the microscopic work. As
we can see from table \ref{tab:The-energy-gain} we find a substantial
effect of about $45\,\text{meV}$, in reasonable agreement with the
more accurate calculations above. All of this indicates the effect
of the polar interactions; it probably misses some of the additional
hydrogenic interactions with graphene seen in the calculations.

\section{Results}

\begin{table}
\caption{Energy per molecule, compared to the lowest energy solution for those
configurations within $50\text{ meV}$ of the ground state. The accuracy
of the energies is about 1 meV, for the stresses about $0.02\text{ GPa}$.
We compare three values of the confinement width (modeled by using
a 3-9 potential with appropriate strength).\label{tab:lowd}}

\resizebox{\textwidth}{!}{%
$
\begin{array}{ccccccccccccc}
\text{configuration } & \alpha= & 350\text{meV/\ensuremath{\text{Å}^{2}}} &  &  & \alpha= & 200\text{meV/\ensuremath{\text{Å}^{2}}} &  &  & \alpha= & 50\text{meV/\ensuremath{\text{Å}^{2}}}\\
 & \Delta E & \Delta E_{\text{conf }} & S_{x,y} & S_{z} & \Delta E & \Delta E_{\text{conf }} & S_{x,y} & S_{z} & \Delta E & \Delta E_{\text{conf }} & S_{x,y} & S_{z}\\
 & \text{meV} & \text{meV} & \text{GPa} & \text{GPa} & \text{meV} & \text{meV} & \text{GPa} & \text{GPa} & \text{meV} & \text{meV} & \text{GPa} & \text{GPa}\\
\text{corrugated I} & 44.6 & 7.2 & 0.6 & 0.03 & 16.4 & 43.9 & 0.16 & 0.13 & -69.3 & 29.0 & -0.12 & 0.1\\
\text{corrugated II} & 30.0 & 4.3 & 0.5 & 0.04 & -18.8 & 36.4 & -0.04 & 0.13 & 32.2 & 52.9 & 0.36 & 0.09\\
\text{corrugated III} & 15.0 & 30.2 & 0.29 & 0.1 & 9.9 & 36.4 & 0.14 & 0.13 & -70.4 & 26.1 & -0.11 & 0.12\\
\text{corrugated IV} & 20.6 & 37.1 & 0.28 & 0.1 & 11.7 & 42.1 & 0.13 & 0.12 & -72.5 & 28.4 & -0.13 & 0.1\\
\text{corrugated V} & 0 & 18.2 & 0.28 & 0.06 & 1.7 & 23.7 & 0.17 & 0.1 & -70.6 & 20.2 & -0.07 & 0.11\\
\text{corrugated VI} & 0.1 & 15.6 & 0.31 & 0.03 & 0 & 30.9 & 0.15 & 0.08 & -75.8 & 25.3 & -0.12 & 0.08\\
\text{in-plane I} & 123. & 0 & 0.82 & 0.01 & 137. & 0 & 0.82 & 0.01 & 104 & 0 & 0.82 & 0.02\\
\text{in-plane II} & 81.0 & 0 & 0.59 & 0. & 95. & 0 & 0.6 & 0. & 62.2 & 0 & 0.6 & 0.01\\
\text{in-plane III} & 21.6 & 0.2 & 0.5 & 0.01 & 34.5 & 0.4 & 0.5 & 0.02 & -54.7 & 28.0 & -0.03 & 0.12\\
\text{in-plane IV} & 81.0 & 0 & 0.54 & 0.02 & 14.8 & 0.9 & 0.4 & 0.03 & 62.2 & 0 & 0.55 & 0.01\\
\text{in-plane V} & 5.2 & 0 & 0.54 & -0.01 & 19.0 & 0 & 0.53 & -0.01 & -75.8 & 25.4 & -0.12 & 0.09\\
\text{out of plane square} & 34.5 & 15.6 & 0.41 & 0.07 & 41.7 & 9.4 & 0.44 & 0.06 & 0 & 3.4 & 0.39 & 0.02
\end{array}
$
}
\end{table}
We now perform a first principle study of water confined between graphene
using density-functional theory (DFT). We immediately need to face
two problems: which of the many functionals to choose, and how to
model the confinement. Water is a notoriously difficult system to
describe using DFT, as reviewed by Gillan \emph{et al.}
\cite{gillan2016perspective}.
All good simulations require van-der-Waals exchange functionals, and
for layered systems and water a combination of van-der-Waals exchange
with optimized exchange potentials appears to be important. For that
reason, and the fact that our simulations are performed with quantum
espresso\cite{giannozzi2009quantum}, we shall compare results using
two functionals, vdW-DF \cite{dion2004vander,thonhauser2007vander,roman-perez2009efficient,sabatini2012structural}
and compare to numbers from vdW-DF-ob86\cite{klimevs2011vander}. 
In each case confinement
is implemented as an external 3-9 potential acting on the O nuclei.
Its effect is actually surprisingly small on the structures--but of
course that may be different for their stability. The most important
difference between our work and previous work that looked at the faces
of confined water is that we work at constant density, and minimize
the energy with respect to the shape of the unit cell, 
rather than an artificially designed 2D enthalpy. As other
calculations, we work at zero temperature.

\begin{figure}
\begin{centering}
\subfloat[corrugated III]{\begin{centering}
\begin{tabular}{|c|}
\hline 
\includegraphics[width=4cm]{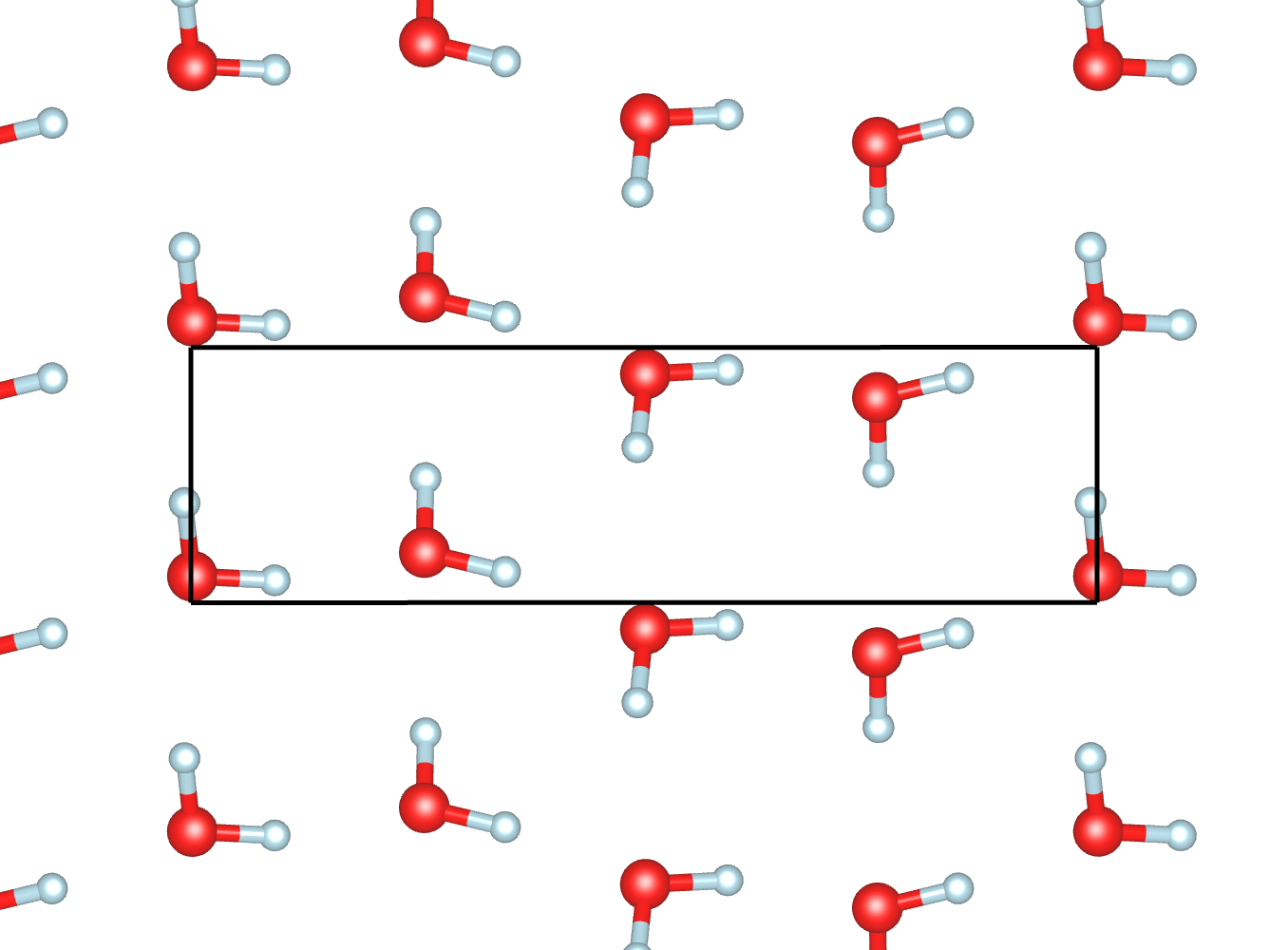}\tabularnewline
\hline 
\includegraphics[width=4cm]{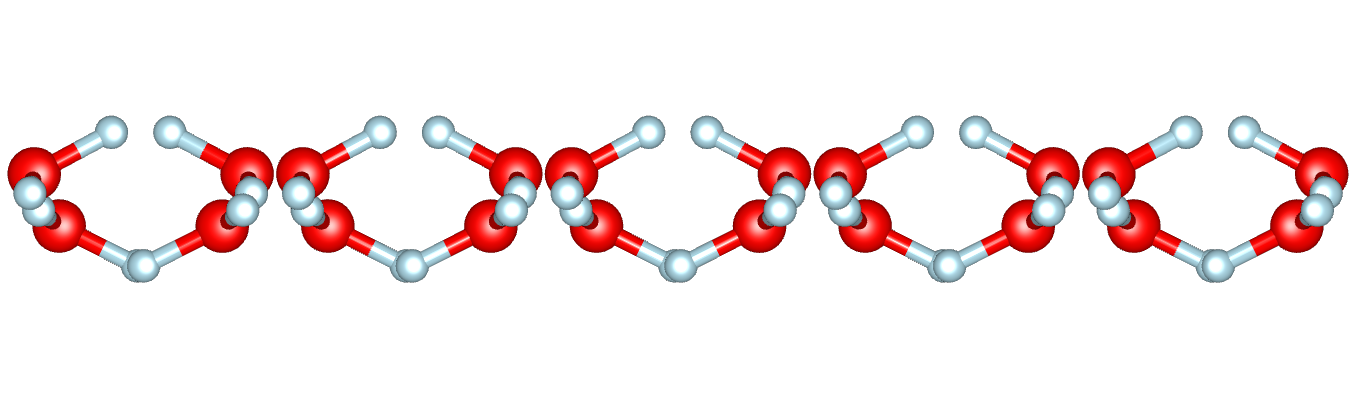}\tabularnewline
\hline 
\end{tabular}
\par\end{centering}

}\subfloat[corrugated V]{\begin{centering}
\begin{tabular}{|c|}
\hline 
\includegraphics[width=4cm]{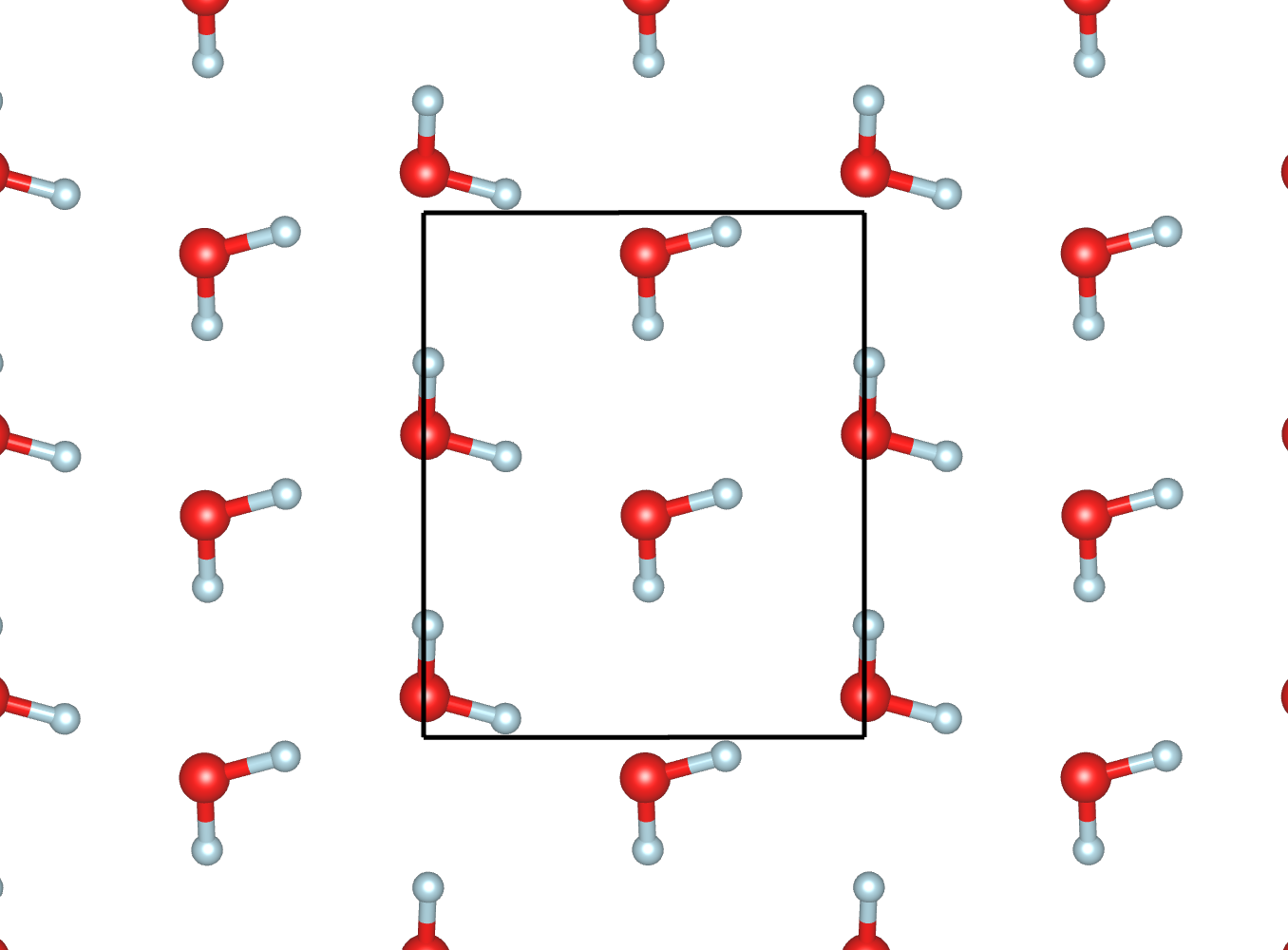}\tabularnewline
\hline 
\includegraphics[width=4cm]{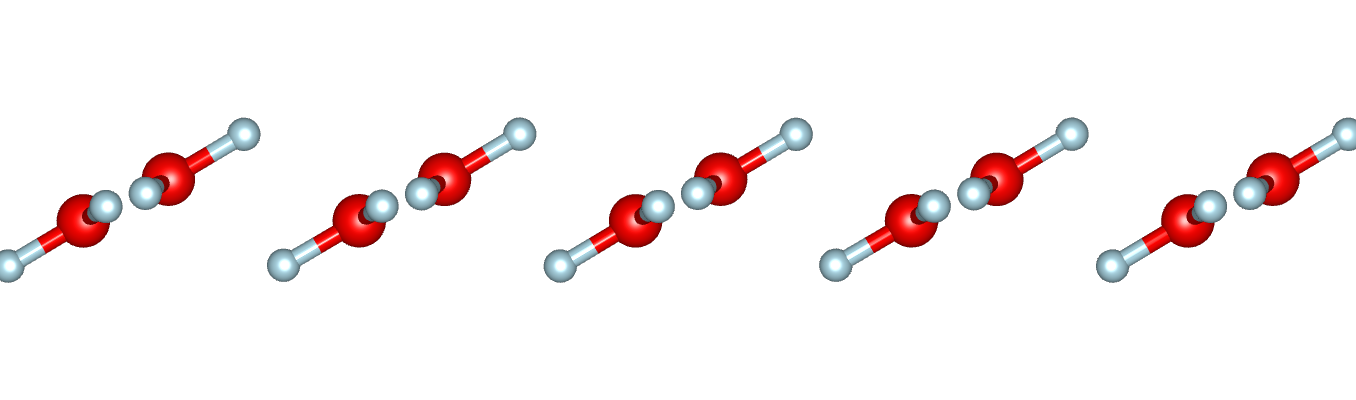}\tabularnewline
\hline 
\end{tabular}
\par\end{centering}

}\subfloat[corrugated VI]{\begin{centering}
\begin{tabular}{|c|}
\hline 
\includegraphics[width=4cm]{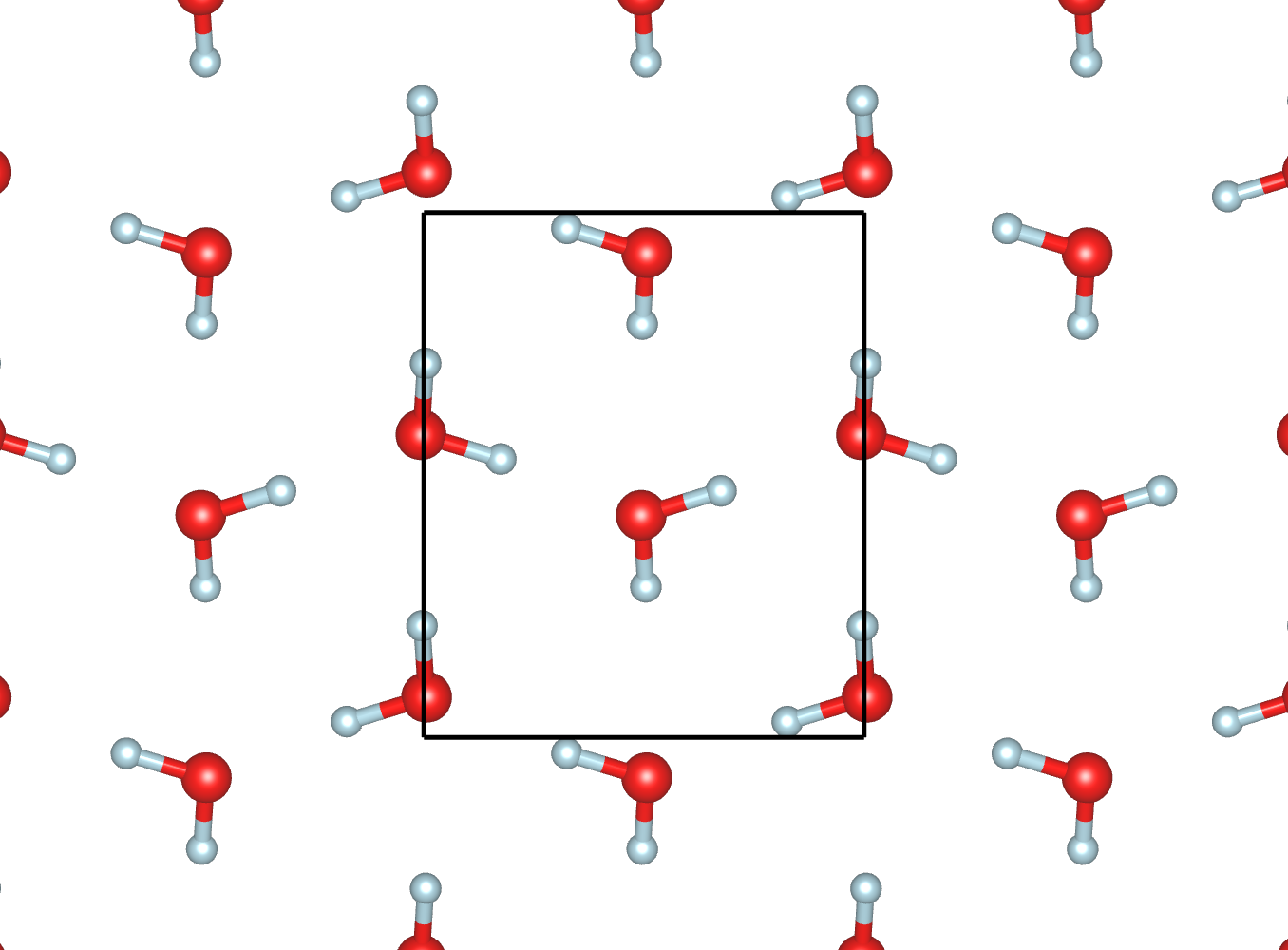}\tabularnewline
\hline 
\includegraphics[width=4cm]{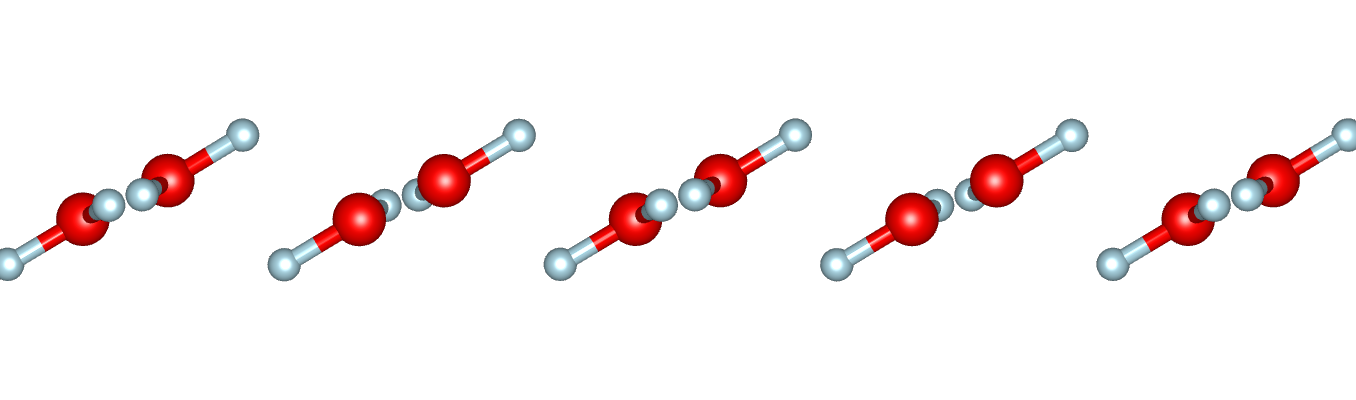}\tabularnewline
\hline 
\end{tabular}
\par\end{centering}

}
\par\end{centering}

\centering{}\subfloat[in-plane V]{\begin{centering}
\begin{tabular}{|c|}
\hline 
\includegraphics[width=4cm]{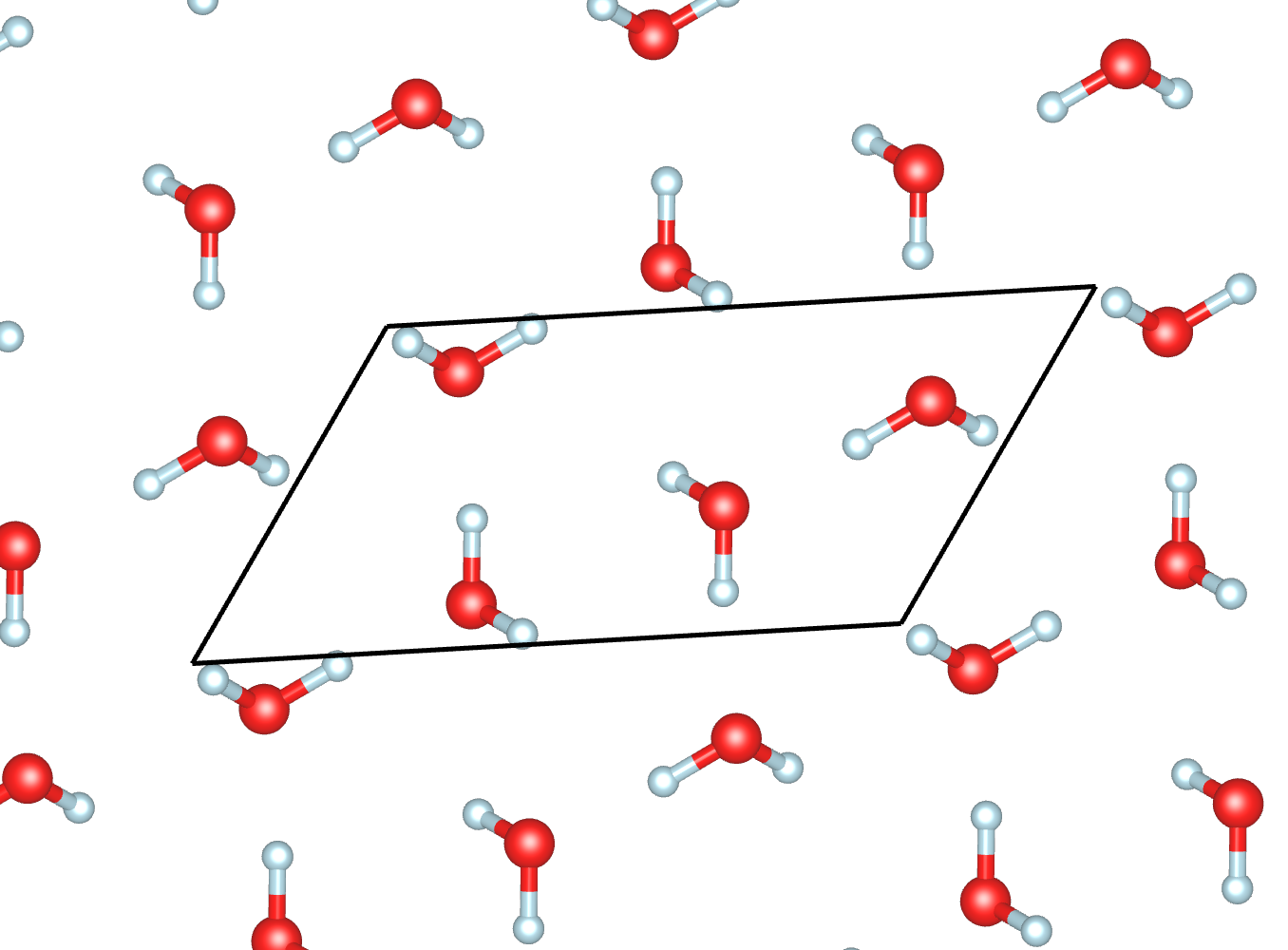}\tabularnewline
\hline 
\includegraphics[width=4cm]{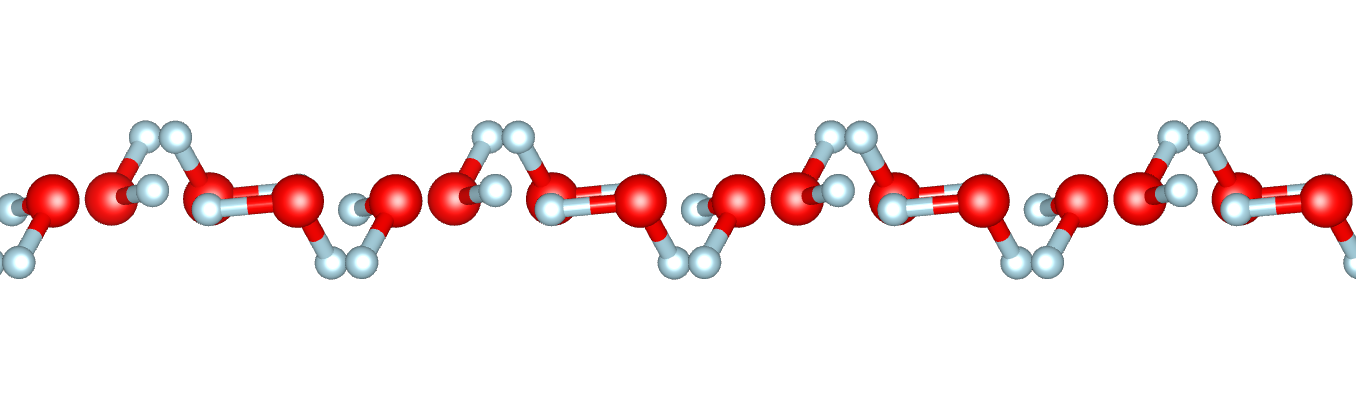}\tabularnewline
\hline 
\end{tabular}
\par\end{centering}

}\subfloat[out of plane square]{\begin{centering}
\begin{tabular}{|c|}
\hline 
\includegraphics[width=4cm]{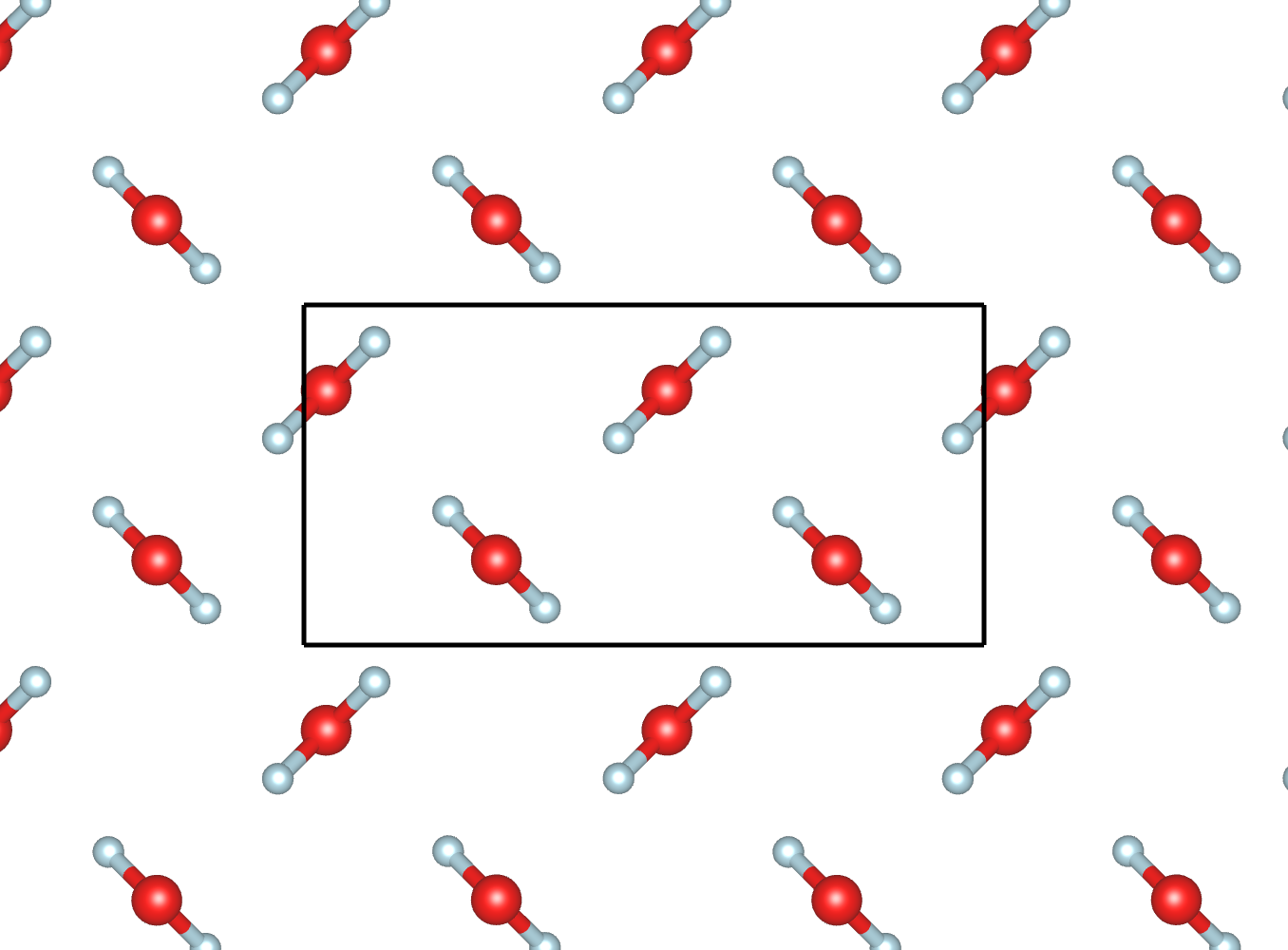}\tabularnewline
\hline 
\includegraphics[width=4cm]{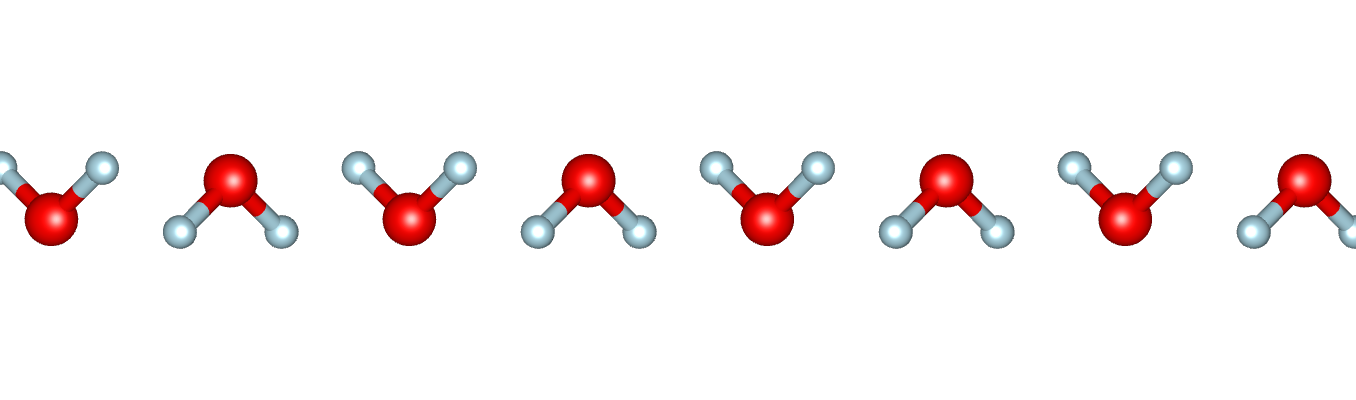}\tabularnewline
\hline 
\end{tabular}
\par\end{centering}

}\caption{A summary of the most relevant configurations.
They are shown here for the strongest confinement, $\alpha=350\text{meV/\ensuremath{\text{Å}^{2}}}$;
the area delimited by the black lines shows the optimized 2D unit
cell.\label{fig:A-few-important}}
\end{figure}

We have used a structure search which, rather than completely random,
is based on the configurations found in the literature\footnote{We have also tried some random searches, but have not found any states
other than the out-of-plane square.}. These correspond to states with all the oxygen atoms
in a single plane, or ones where the oxygen atoms form a slightly corrugate plane. 
Additional detail
can be found in the supplementary material. Considering the potential orientational energy gains, one must consider
all configurations where the energy per molecule is less than $\simeq50\text{ meV}$,
and then weigh which ones may be favored. We use results from
the literature
\cite{chen2016twodimensional,corsetti2016structural} to select potentially
appropriate candidates,\footnote{We have also tried some random searches, but have not found any states
other than the out-of-plane square.} and then minimize the energy of such states,
first with respect to position in the unit cell, and then with respect
to the shape of the unit cell, keeping the density fixed. Stable minimum
energy solutions for any given density then have isotropic 2D stress.
Labeling of the configurations is with the initial configurations--see
the references for details. Numerical details of the results can be
found in table \ref{tab:lowd}.

We show some of the key configurations in figure~\ref{fig:A-few-important},
where we selected the strong confining potential; the main visual
difference for weaker confinement is a further separation of the ``square''
configurations in a bilayer like structure. The lowest energy states
(for this confinement) both have a rhombic unit cell--in this case
it looks like there is a perfect equilateral triangular crystal, but
with rather different dipole orientations. The real surprise is the
perfectly square crystal state, which seem relatively insensitive
to the detail of the confining potential.

The water molecules experience very little confinement in the lowest
energy solutions, and the vertical stresses (indicative of how hard
the water pushes on the graphene, separating the layers) are 
small, as can be seen from Table \ref{tab:lowd}. In that table, we
label the states with the initial seeding configuration--which in
some cases is close to the final result. This is the case where we
want to concentrate most clearly on the effect of the confining potential.
All of these are single layers, with some corrugation, but even for
a weak confining potential we find that all oxygens are located very close to
$d/2$. We can thus investigate the effect of moderate changes
in the confining potential by just look at the harmonic potential
dependence around the midpoint, $E=\alpha\delta r^{2}$. We take
the value obtained from the $3-9$ potential ($\alpha=350\text{meV}/\text{Å}^{2}$), which 
is probably unrealistically steep
about the minimum due to the very strong repulsive core, as an upper
estimate. We will use  $\alpha=50$, $200$ and $350\text{meV}/\text{Å}^{2}$
to investigate this dependence.

As we can see in table \ref{tab:lowd}, the ordering of the states and the stresses are
rather sensitive to the choice of this parameter. For the strongest confinement, we
find that those configurations relaxed from a slightly out of plane
rhombic unit cell give by far the lowest energies, but in some cases
with substantial vertical stresses. As we decrease the confinement,
the vertical stress gets larger, corresponding to a tendency towards
forming bilayers, and for the lowest harmonic strength we find a tendency
to form bilayers--which prefer a slightly higher density inside each
layer, thus leading to negative stress. For the flatter honeycomb
based states the trend is that we need a slightly larger horizontal
stress to confine the water, maybe slightly closer to what we expect
the stress to be, vertical stresses are negligible. As we decrease
the confinement, some of these states also start forming bilayers.
As we decrease the confinement, the only really square configuration
we have found, the anti-ferroelectric states with the dipoles are
perpendicular to the graphene, becomes the lowest energy stable state,
thus giving at least a natural candidate for a real square crystal.

Finally, we look at the orientation dependent potential Eq.~(\ref{eq:VHVO}).
For this case find that the solutions that tend to form a bilayer
now have substantial vertical stresses, and are thus probably unstable
against an increase in separation of the graphene layers, see table
\ref{tab:lowd-1}. The anti-ferroelectric state gains considerable
energy relative to the in-plane solutions, and occurs about $16\,\text{meV}$
above the lowest state. This is very suggestive that this definitely
needs to be considered as the prime candidate for the ground state,
since minor modification to the confining potential can bring its
energy down considerably.

\begin{table}
\caption{Energy per molecule, compared to the lowest energy solution for those
configurations near the ground state. The accuracy of the energies
is about 1 meV, for the stresses about $0.02\text{ GPa}$. We compare
a tight directionless confinement with a directional potential of
the same depth and similar range (modeled by a combination of 3-9
potentials with appropriate strengths).\label{tab:lowd-1}}

\[
\begin{array}{ccccccccc}
\text{configuration} & \alpha= & 350\text{meV/\ensuremath{\text{Å}^{2}}} &  &  & \text{directional}\\
 & \Delta E & \Delta E_{\text{conf }} & S_{x,y} & S_{z} & \Delta E & \Delta E_{\text{conf }} & S_{x,y} & S_{z}\\
 & \text{meV} & \text{meV} & \text{GPa} & \text{GPa} & \text{meV} & \text{meV} & \text{GPa} & \text{GPa}\\
\text{corrugated I} & 44.6 & 7.18 & 0.6 & 0.03 & 18.2 & 37.0 & 0.22 & 0.16\\
\text{corrugated II} & 30.0 & 4.29 & 0.5 & 0.04 & 10.7 & 30.6 & 0.20 & 0.16\\
\text{corrugated III} & 15.0 & 30.2 & 0.29 & 0.10 & 10.7 & 30.9 & 0.20 & 0.16\\
\text{corrugated IV} & 20.6 & 37.1 & 0.28 & 0.10 & 11.3 & 34.0 & 0.17 & 0.14\\
\text{corrugated V} & 0 & 18.2 & 0.28 & 0.06 & 1.64 & 19.3 & 0.22 & 0.13\\
\text{corrugated VI} & 0.1 & 15.6 & 0.31 & 0.03 & 0 & 26.4 & 0.2 & 0.11\\
\text{in-plane I} & 123. & 0 & 0.82 & 0.01 & 104. & 4.16 & 0.7 & 0.07\\
\text{in-plane II} & 80.9 & 0 & 0.59 & 0. & 96.9 & 4.24 & 0.61 & 0.08\\
\text{in-plane III} & 21.6 & 0.233 & 0.5 & 0.01 & 29.7 & 7.33 & 0.44 & 0.06\\
\text{in-plane IV} & 81.0 & 0 & 0.54 & 0.02 & 96.8 & 4.28 & 0.55 & 0.06\\
\text{in-plane V} & 5.16 & 0 & 0.54 & -0.01 & 17.2 & 0 & 0.54 & 0.02\\
\text{out of plane square} & 34.5 & 15.6 & 0.41 & 0.07 & 33.3 & 3.74 & 0.42 & 0.04
\end{array}
\]
\end{table}

Finally, we still need to consider whether the use of a different
van-der-Waals exchange functional might change the results. We have
repeated the calculation for the modified DFT function vdW-DF-ob86,
and find that energy shifts by a few meV. Such changes are small compared
to the uncertainties in the nature of the confinement, and can thus
be safely ignored for now.

\section{Conclusions}

We have analyzed some of the effect of confinement on the phases
of 2D water. We find that there is a great sensitivity to the nature
of the confining potential, but we also see a perfectly good candidate
for the observed square crystal. Clearly it is of crucial importance
to understand the nature of the confinement experienced of water.
It is important in water flow experiments\cite{radha2016molecular},
and when using graphene anvils \emph{\cite{vasu2016vander}.}. The
next stage of this work will be to get a better understanding of
the water graphene interactions from a combination of macroscopic and
microscopic methods.

\section{methods}
\paragraph{Computational Technology}
All DFT calculation were performed using the Quantum Espresso
code\cite{giannozzi2009quantum}, version 5.1.2, with modifications to
account for the confining potentials.  The geometry optimizations were
all performed using the vdW-DF exchange functional
\cite{dion2004vander,thonhauser2007vander,roman-perez2009efficient,sabatini2012structural},
with  Perdew-Burke-Ernzerhof (PBE) pseudo potentials. The kinetic
energy cutoff was chosen to be 450 eV.

\begin{acknowledgement}
The author would like to thank Paco Guinea and Mike Moore for
useful discussions regarding the work reported in this paper, and
Paco Guinea for sharing his notes on bubble formation. The author would
also like to acknowledge the assistance given by IT Services and the
use of the Computational Shared Facility at The University of Manchester,
where most of the calculations were performed.
\end{acknowledgement}

\begin{suppinfo}
A full list of all configurations is given in the supplementary materials, where we also discuss
the starting values used.
\end{suppinfo}

\bibliography{squareice}

\end{document}